\documentclass[]{spie}  

\usepackage{subcaption}
\usepackage{amsmath,amsfonts,amssymb}
\usepackage{graphicx}
\usepackage{wrapfig}
\usepackage{caption}
\usepackage{booktabs}
\usepackage{float}
\floatstyle{plaintop}
\restylefloat{table}
\usepackage{float}
\usepackage{blindtext}
\usepackage{multicol}
\usepackage{tabularx}
\usepackage[colorlinks=true, allcolors=blue]{hyperref}
\newcolumntype{P}[1]{>{\centering\arraybackslash}p{#1}}

\title{Automatic tumour segmentation in H\&E-stained whole-slide images of the pancreas.}

\author[a]{P. Vendittelli}
\author[a]{E.M.M. Smeets}
\author[a]{G.J.S. Litjens}
\affil[a]{Department of Pathology, Radboud University Medical Center, Nijmegen, The Netherlands}

\begin{document} 
\maketitle

\begin{abstract}
Pancreatic cancer will soon be the second leading cause of cancer-related death in Western society. Imaging techniques such as CT, MRI and ultrasound typically help providing the initial diagnosis, but histopathological assessment is still the gold standard for final confirmation of disease presence and prognosis. In recent years machine learning approaches and pathomics pipelines have shown potential in improving diagnostics and prognostics in other cancerous entities, such as breast and prostate cancer. A crucial first step in these pipelines is typically identification and segmentation of the tumour area. Ideally this step is done automatically to prevent time consuming manual annotation. In this paper, we propose a multi-task convolutional neural network to balance disease detection and segmentation accuracy. We validated our approach on a dataset of 29 patients (for a total of 58 slides) at different resolutions. The best single task segmentation network achieved a median Dice of 0.885 (0.122) IQR at a resolution of 15.56 $\mu$m. Our multi-task network improved on that with a median Dice score of 0.934 (0.077) IQR. 
\end{abstract}


\section{INTRODUCTION}
\label{sec:intro}  

Pancreatic ductal adenocarcinoma (PDAC) is now the seventh leading cause of cancer-related deaths worldwide \cite{rawla2019epidemiology} and will soon become the second leading cause of cancer-related death in Western society \cite{mcguigan2018pancreatic}.  Europe has the highest burden of PDAC in the world, with 150 000 new cases in 2018 and 95 000 deaths/year. Worldwide there were over 500 000 deaths in 2020\cite{sung2021global}. Moreover, PDAC has the lowest survival of all types of cancers \cite{siegel2020cancer} (median survival time 4.6 months), with patients losing 98\% of their healthy life expectancy.

Although imaging techniques as computed tomography (CT), magnetic resonance imaging (MRI) and endoscopic ultrasound (EUS) \cite{mcguigan2018pancreatic} are widely used for an initial diagnosis of pancreatic cancer, biopsies are the gold standard for final confirmation of disease presence and prognosis. In recent years machine learning approaches and pathomics pipelines have shown potential in improving diagnostics and prognostics in other cancerous entities, such as breast and prostate cancer \cite{van2021hooknet}\cite{Bulten_2020}, but very little research has been done on the analysis of pancreatic histopathological images through automatic systems. As far as the authors are aware, Fu et al.\cite{fu2021automatic} proposed an automatic method based on convolutional neural networks (CNNs) for pancreatic cancer detection and segmentation. They perform patch- and Whole-slide Image- level classification to determine which areas of an image are most critical for PDAC identification and use the segmentation network as control study for their classification network. A key challenge when combining detection and segmentation in a single convolutional neural network is to balance detection and segmentation accuracy. This paper proposes a  multi-task convolutional neural network for segmenting PDAC on whole-slide images (WSIs) and compares it to a single-task segmentation network. Furthermore, we explore which image resolution is most suitable for PDAC detection and segmentation.

\section{METHODOLOGY}
 
\begin{figure} [t!]
    \centering
    \includegraphics[height=8.3cm]{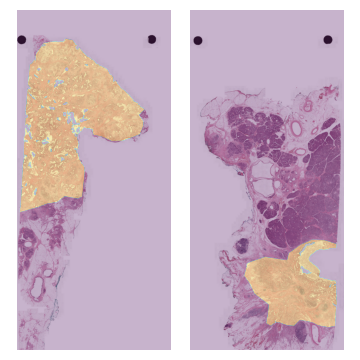}
    \caption{Example of two cases with coarse tumour annotations, masked with tissue masks. Note that the annotations include both stromal cells and tumour cells.}
    \label{fig:example_data}
\end{figure}

\paragraph{Data}
For this study, 29 patients were selected from a PET-cohort of 158 patients with histologically proven pancreatic ductal adenocarcinoma (PDAC) who underwent [18F]FDG PET/CT in the period between 2004 and 2015 at Radboud University Medical Center, Nijmegen (The Netherlands) during diagnostic work-up for a total of 58 (2 slides per patient)  H\&E stained slides. Fifty-one slides contained cancerous tissue, while seven only had normal tissue. All the positive cases were collected with tumour annotations (Figure \ref{fig:example_data}) made under the supervision of a pathologist. Additionally, the annotations were masked with the tissue masks generated by a tissue-background segmentation algorithm\cite{bandi2019resolution} in order to remove possible presence of the background. Slides were organised in 5 folds for cross-validation, maintaining a ratio for training  - validation of 0.8 - 0.2. There were no slides coming from the same patient in both the training and validation folds. Due to the limited data available, we did not use a separate test fold.
 
\label{sec:title}

\paragraph{Network architecture and experiment setup}
WSIs have a typical dimension of $\sim$ 100,000 x $\sim$ 100,000 pixels at the highest resolution, around 0.25 $\mu$m. Forwarding even a single slide through any CNN would be nearly impossible for modern GPU as it would require a computational power and memory higher than any of the current hardware, unless using specialized end-to-end approaches\cite{pinckaers2019streaming} \cite{8809829}. For segmentation, a patch-based approach is typically preferred due to its simplicity and because generally only local patch-level information is required to solve the task.

The best resolution at which to sample patches is often unknown a-priori, thus in our study, we compared the performances of the same model architecture over three different patch sets sampled at three different resolutions, respectively: 3.89$\mu$m, 7.78$\mu$m and 15.56$\mu$m. Patches of shape 256x256 pixels were sampled with equal probability from normal tissue and cancerous tissue in both the  training and validation sets. We augmented the sampled training patches for every experiment by applying flipping (both horizontal and vertical), blurring, modification of HSV channels, contrast and brightness.

We used a U-Net\cite{ronneberger2015u} model provided by the open-source Segmentation library\cite{Yakubovskiy:2019} as our segmentation model. We selected a five-layers efficientnet-b0 structure, pre-trained on ImageNet, as the backbone for the encoder and a five-layers decoder with skip-connection for reconstructing the output. For each experiment, we trained the U-Net for 30 epochs and used Adam optimizer (learning rate =$1e-5$) with a mini-batch of 16 patches. A Dice loss was used during training. During inference, we divide the test WSIs into sequential patches sampled at the exact resolution the model was trained and pass it through the network to obtain the cancer probability map then binarized with a threshold of 0.5 for generating the  segmentation of the tumour.

For the multi-task network, we added a single neuron acting as a binary classifier to the last layer (latent space) of the encode.  For each patch passed through the network, there would then be two outputs: a segmentation mask and a label. To assign a ground truth label for the patch, we tested two different criteria: in the \textbf{first} (Seg + Cl1), the patch is given label 1 (cancer) if at least 51\% of its pixels belong to the tumour class, while in the \textbf{latter} (Seg +Cl2) a patch is considered cancer if there is at least 1 pixel belonging to the tumour class in it. \newline To optimize this small binary classifier we used the Binary Cross-Entropy (BCE) loss. Thus, the total loss function used for training the multi-task network was the arithmetic sum of the segmentation loss and the classification loss, therefore the Dice loss function and the Binary Cross-Entropy loss. As reported in Table \ref{tab:Dices}, the single-task network achieved higher median dice when trained at lower resolution, therefore we trained the multi-task network with samples extracted at 15.56$\mu$m and for 100 epochs. The reason why the previous experiments were done with 30 epochs of training is that we noticed a plateau after that amount of epochs, while adding this classification task allowed us to train the network longer without overfitting.
During inference, we rely on the classification layer to produce the output mask. If a patch is classified as positive, so containing cancer, then the product of the segmentation branch is kept, while if a patch is classified as negative, the result of the segmentation branch is masked with an array full of zeros. This ideally should lead to a reduction of false positive patches.   

\paragraph{Results}
\begin{figure}[ht!]
\centering
  \includegraphics[width=0.6\textwidth]{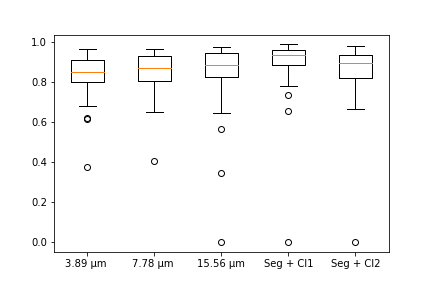}
  \caption{Boxplots of the network over the three different resolutions and multi-task network on the positive cases.}
  \label{fig:boxplots}
\end{figure}

\begin{table}
	\caption{Median and Median positive Dice for the three resolutions and over the Segmentation + Classification Task.\label{tab:Dices}}
	\vspace*{1mm}
		\centering
		 \begin{tabular}{|l|l|l|}
\hline
Experiment                           & Med. Dice  (IQR)      & Med. Pos. Dice  (IQR)      \\ \hline
3.89$\mu$m                           & 0.831  (0.151)               & 0.846  (0.110)                  \\ \hline
7.78$\mu$m                           & 0.848  (0.149)               & 0.867  (0.124)                     \\ \hline
15.56$\mu$m                          & 0.874  (0.160)               & 0.885  (0.122)                     \\ \hline
\textbf{Seg + Cl 1}                           & \textbf{0.926  (0.143)}              &\textbf{0.934  (0.077)}                     \\ \hline
Seg + Cl 2                           & 0.871  (0.129)               & 0.893  (0.117)                     \\ \hline
\end{tabular}

\end{table}

We evaluated the performances of our models calculating the Dice coefficient on the validation images per each fold. Figure \ref{fig:boxplots} shows the boxplots of the experiments over the five-fold cross-validation. When lowering the resolution at which to extract patches, the model has a higher median Dice coefficient for segmenting the tumour. This might be due to the fact that for pancreatic cancer segmentation architectural features of the whole tissue are more important than cellular features. Therefore, to achieve good performance is not needed to increase the resolution. Table \ref{tab:Dices} reports the median Dice coefficient with inter-quartile range calculated for the whole dataset (58 cases) and the positive cases only (51). From the table is possible to see that adding the classification branch to the network does improve the results, but this is dependant on the labeling criteria assigned. Indeed, with the second labeling criteria (Seg + Cl 2) nearly no additional information is given to the single-task network, and this might be due to the fact that this might be a really permissive labeling criteria, as a patch is given label 1 if there's at least one pixel belonging to the tumour class in its ground truth patch. Thus, results are comparable with the single-task network, while on the other hand, when using the first labeling criteria (Seg + Cl 1), the network actually feels the effect of the more restrictive criteria. the  Figure \ref{fig:boxplots} shows some outliers with Dice of 0. Most of them are the negative cases, for which the model generated false positive patches, but for the multi-task network and the lower resolution approach, despite having the overall best results there is also a missed positive case, for which the segmentation mask was negative, therefore for that case as well the Dice was set to 0.  
\newline
\paragraph{Discussion}
\begin{figure}[t!]
  \includegraphics[width=\textwidth,height=8.3cm]{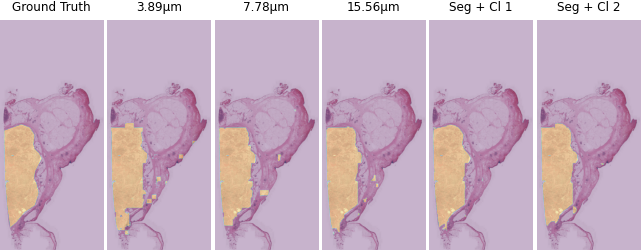}
  \caption{Example case containing PDAC. Tumor segmentation is overlayed in yellow. 3.89$\mu$m (Dice: 0.926), 7.78$\mu$m (Dice: 0.960), 15.56$\mu$m (Dice: 0.959), Seg+Cl1 (Dice: 0.978), Seg+Cl2 (Dice: 0.967)}
  \label{fig:results_a}
\end{figure}
\begin{figure}[ht]
  \includegraphics[width=\textwidth, height=8.3cm]{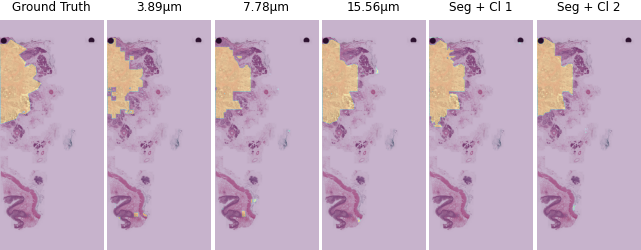}
  \caption{Example case containing PDAC. Tumor segmentation is overlayed in yellow. 3.89$\mu$m (Dice: 0.823), 7.78$\mu$m (Dice: 0.849), 15.56$\mu$m (Dice: 0.939), Seg+Cl1 (Dice: 0.947), Seg+Cl2 (Dice: 0.901)}
  \label{fig:results_b}
\end{figure}

\begin{figure}[ht]
  \includegraphics[width=\textwidth,height=8.3cm]{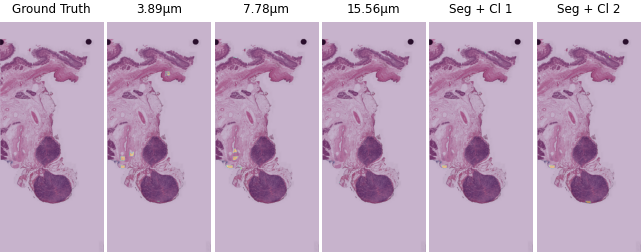}
  \caption{Example case without PDAC. The lower resolution networks show fewer false positives both for the single- and multi-task networks. Where small false positive patches are generated, the Dice is set to 0.}
  \label{fig:falsepos_a}
\end{figure}

\begin{figure}[ht]
  \includegraphics[width=\textwidth,height=8.3cm]{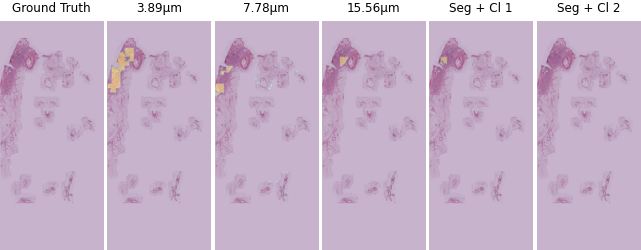}
  \caption{Example case without PDAC. The lower resolution networks show fewer false positives both for the single- and multi-task networks. Where small false positive patches are generated, the Dice is set to 0.}
  \label{fig:falsepos_b}
\end{figure}

Figure \ref{fig:results_a} and Figure \ref{fig:results_b} compare the segmentation masks produced on two positive cases by the model trained on the three resolutions and the model trained on the lower resolution in the multi-task settings. For the first case, the models obtained a high Dice score for all the experiments, whereas in the second case the models obtained a Dice score within the inter-quartile range. 
Figure \ref{fig:results_a} and Figure \ref{fig:results_b} show, the lower the resolution, the better the network can segment the edges of the tumour. This is because at higher resolutions, the patch contains less spatial information; thus, the model would nearly-always deal with patches completely positive or completely negative, while at lower resolution this happens less frequently, resulting in the edges of the segmented tumour being better defined. The network also is forced to focus more on architectural aspects of the tissue instead of cellular features, which might be more important in PDAC. The addition of the multi-task loss shows a further improvement in median Dice and a reduction in inter-quartile range. The multi-task network and the network trained at lower resolution miss classify a single positive case. This might due to the composition of the tissue itself which looks very different from the others in the dataset, thus at higher resolution it was miss classified, while detected (with false positive patches) at higher resolution.

In PDAC-negative cases the networks trained at higher resolutions usually produce more false-positive patches, while training the network at lower resolution decreases the number of false-positive patches (Figure \ref{fig:falsepos_a} and Figure \ref{fig:falsepos_b}). In the multi-task setting this is even further improved. In general, the percentage of miss-classified tissue for the seven negative cases in our dataset varies from 3.2$\pm$ 1.66\% for the model trained at 3.89$\mu$m to 0.60 $\pm$ 0.57\% for the best multi-task configuration. 
\newline This work can be considered as the first step for more detailed tissue analysis. Once the main tumour area is accurately segmented, discrimination on the tumour microenviroenment (TME) components, such as epithelial cells and stromal cells, can be done, yielding to possible prognostic markers such as Tumour to stroma ratio (TSR).\newline Our work does have several limitations. Due to the small dataset size we needed to perform five-fold cross-validation to adequately train the networks. In future work we want to expand the dataset, include data from external sources and analyse whether the improvement provided by the multi-task network is statistically significant. Additionally, we did not explore more sophisticated classification branches for the multi-task network or different weights for the classification loss, which could help further improve performance. Last, due to the fact that we only have coarse tumor annotation we cannot separate between different tumour components, such as stroma, immune cells and tumor cells, which will be important for extracting relevant pathomics features. Summarizing, we have shown that standard segmentation architectures can segment PDAC accurately, even at low-resolution. Furthermore, in a multi-task setting performance is further improved.

\acknowledgments
This project has received funding from the European Union’s Horizon 2020 research and innovation programme under grant agreement No 101016851, project PANCAIM.

\bibliography{report} 

\begin{thebibliography}{10}

\bibitem{rawla2019epidemiology}
Rawla, P., Sunkara, T., and Gaduputi, V., ``Epidemiology of pancreatic cancer:
  global trends, etiology and risk factors,'' {\em World journal of
  oncology}~{\bf 10}(1),  10 (2019).

\bibitem{mcguigan2018pancreatic}
McGuigan, A., Kelly, P., Turkington, R.~C., Jones, C., Coleman, H.~G., and
  McCain, R.~S., ``Pancreatic cancer: A review of clinical diagnosis,
  epidemiology, treatment and outcomes,'' {\em World journal of
  gastroenterology}~{\bf 24}(43),  4846 (2018).

\bibitem{sung2021global}
Sung, H., Ferlay, J., Siegel, R.~L., Laversanne, M., Soerjomataram, I., Jemal,
  A., and Bray, F., ``Global cancer statistics 2020: Globocan estimates of
  incidence and mortality worldwide for 36 cancers in 185 countries,'' {\em CA:
  a cancer journal for clinicians}~{\bf 71}(3),  209--249 (2021).

\bibitem{siegel2020cancer}
Siegel, R.~L., Miller, K.~D., and Jemal, A., ``Cancer statistics, 2020,'' {\em
  Ca-a Cancer Journal for Clinicians}~{\bf 70}(1),  7--30 (2020).

\bibitem{van2021hooknet}
van Rijthoven, M., Balkenhol, M., Sili{\c{n}}a, K., van~der Laak, J., and
  Ciompi, F., ``Hooknet: Multi-resolution convolutional neural networks for
  semantic segmentation in histopathology whole-slide images,'' {\em Medical
  Image Analysis}~{\bf 68},  101890 (2021).

\bibitem{Bulten_2020}
Bulten, W., Pinckaers, H., van Boven, H., Vink, R., de~Bel, T., van Ginneken,
  B., van~der Laak, J., Hulsbergen-van~de Kaa, C., and Litjens, G., ``Automated
  deep-learning system for gleason grading of prostate cancer using biopsies: a
  diagnostic study,'' {\em The Lancet Oncology}~{\bf 21},  233–241 (Feb
  2020).

\bibitem{fu2021automatic}
Fu, H., Mi, W., Pan, B., Guo, Y., Li, J., Xu, R., Zheng, J., Zou, C., Zhang,
  T., Liang, Z., et~al., ``Automatic pancreatic ductal adenocarcinoma detection
  in whole slide images using deep convolutional neural networks,'' {\em
  Frontiers in oncology}~{\bf 11},  2464 (2021).

\bibitem{bandi2019resolution}
B{\'a}ndi, P., Balkenhol, M., van Ginneken, B., van~der Laak, J., and Litjens,
  G., ``Resolution-agnostic tissue segmentation in whole-slide histopathology
  images with convolutional neural networks,'' {\em PeerJ}~{\bf 7},  e8242
  (2019).

\bibitem{pinckaers2019streaming}
Pinckaers, H., van Ginneken, B., and Litjens, G., ``Streaming convolutional
  neural networks for end-to-end learning with multi-megapixel images,'' {\em
  arXiv preprint arXiv:1911.04432}  (2019).

\bibitem{8809829}
Tellez, D., Litjens, G., van~der Laak, J., and Ciompi, F., ``Neural image
  compression for gigapixel histopathology image analysis,'' {\em IEEE
  Transactions on Pattern Analysis and Machine Intelligence}~{\bf 43}(2),
  567--578 (2021).

\bibitem{ronneberger2015u}
Ronneberger, O., Fischer, P., and Brox, T., ``U-net: Convolutional networks for
  biomedical image segmentation,'' in [{\em International Conference on Medical
  image computing and computer-assisted
  intervention}{\nolinebreak\hspace{0.1em}]},   234--241, Springer (2015).

\bibitem{Yakubovskiy:2019}
Yakubovskiy, P., ``Segmentation models pytorch.''
  \url{https://github.com/qubvel/segmentation_models.pytorch} (2020).

\end{thebibliography}
\bibliographystyle{spiebib} 

\end{document}